\documentclass[preprint,12pt]{elsarticle}

\usepackage{graphicx}
\usepackage{caption}
\usepackage{subcaption}
\usepackage{amsmath}
\usepackage{mathalpha}
\usepackage[colorlinks=true,linkcolor=blue]{hyperref}
\hypersetup{pdfauthor={Name}}
\usepackage{multirow}
\usepackage{array}
\usepackage{float}
\usepackage{placeins}
\newcommand{\PreserveBackslash}[1]{\let\temp=\\#1\let\\=\temp}
\newcolumntype{C}[1]{>{\PreserveBackslash\centering}p{#1}}

\usepackage{amssymb}

\usepackage{lineno}

\biboptions{sort&compress}

\journal{---}

\begin{document}

\begin{frontmatter}

\title{On flame speed enhancement in turbulent premixed hydrogen-air flames during local flame-flame interaction}


\author[fir]{Yuvraj}
\author[fir]{Yazdan Naderzadeh Ardebili}
\author[sec]{Wonsik Song}
\author[sec]{Hong G. Im}
\author[thr]{Chung K. Law}
\author[fir]{Swetaprovo Chaudhuri \corref{cor4}}
\ead{swetaprovo.chaudhuri@utoronto.ca}

\address[fir]{Institute for Aerospace Studies, University of Toronto, Toronto, Canada}
\address[sec]{Clean Combustion Research Center, King Abdullah University of Science and Technology, Thuwal, Saudi Arabia}
\address[thr]{Department of Mechanical and Aerospace Engineering, Princeton University, USA}
\cortext[cor4]{Corresponding author}

\begin{abstract}
Local flame displacement speed $S_d$ of a turbulent premixed flame is of fundamental and practical interest. For H$_2$-air flames, the interest is further accentuated given the recent drive towards the development of zero-carbon combustors for both power and aircraft engine applications. The present study investigates several three-dimensional Direct Numerical Simulation (3D DNS) cases of premixed H$_2$-air turbulent flames to theoretically model the $S_d$ at negative curvatures, building upon recent works. Two of the four DNS cases presented are simulated at atmospheric pressure and two at elevated pressure. The DNS cases at different turbulence Reynolds numbers ($Re_t$) and Karlovitz numbers ($Ka$) are generated using detailed chemistry. It has been shown in the previous studies that at atmospheric pressure, the density-weighted flame displacement speed $\widetilde{S_d}$ is enhanced significantly over its laminar value ($S_L$) at large negative curvature $\kappa$ due to flame-flame interactions. The current work justifiably employs an imploding cylindrical laminar flame configuration to represent the local flame surfaces undergoing flame-flame interaction in a 3D turbulent flame. Therefore, to acquire a deep understanding of the interacting flame dynamics at large negative curvatures, one-dimensional (1D) simulations of an inwardly propagating cylindrical H$_2$-air laminar premixed flame, with detailed chemistry at the corresponding atmospheric and elevated pressure conditions are performed. In particular, the 1D simulations emphasized the transient nature of the flame structure during these interactions. Based on the insights from the 1D simulations, we utilize an analytical approach to model the $\widetilde{S_d}$ at these regions of extreme negative $\kappa$ of the 3D DNS.
The analytical approach is formulated to include the effect of variable density, convection and the inner reaction zone motion. The joint probability density function (JPDF) of $\widetilde{S_d}$ and $\kappa$ and the corresponding conditional averages obtained from 3D DNS showed clear negative correlation between $\widetilde{S_d}$ and $\kappa$ at all pressures.
The obtained model successfully predicts the variation of $\langle\widetilde{S_d}|_{\kappa}\rangle$ with $\kappa$ for the regions on the flame surface with large negative curvature ($\kappa\delta_L \! \ll \! -1$) at atmospheric as well as at elevated pressure, with good accuracy. This showed that the 1D cylindrical, interacting flame model is a fruitful representation of a local flame-flame interaction that persists in a 3D turbulent flame, and is able to capture the intrinsically transient dynamics of the local flame-flame interaction. The 3D DNS cases further showed that even in the non-interacting state at $\kappa=0$, on average $\widetilde{S_d}$ can deviate from $S_L$. $\widetilde{S_d}$ at $\kappa=0$ is a manifestation of the internal flame structure, controlled by turbulence transport in the large $Ka$ regime. Therefore, the correlation of $\langle\widetilde{S_d}\rangle/S_L$ with the the normalized gradient of the progress variable, $\langle|\widehat{\nabla c}|_{c_0}\rangle$ at $\kappa =0$ is explored.

\end{abstract}


\end{frontmatter}


\section{Introduction}
\label{S:1}
The propagation speed of steady, one-dimensional, unstretched, freely propagating laminar flame, $S_L$, is a fundamental quantity determined by thermo-chemical and transport properties of a mixture \cite{law2006}. It emerges as the eigenvalue of the diffusive-reactive transport equations and has been routinely used for validating newly developed chemical reaction mechanisms. In a multi-dimensional flow field, the flame displacement speed, $S_d$, is defined at any point inside the flame structure \cite{echekki1996, Peters2000}, and represents the local velocity of the flame surface (with position vector $\boldsymbol{x}_F$) relative to the local fluid velocity $\boldsymbol{u}$, along the local unit normal vector $\boldsymbol{n}$ 
to the flame surface, where $\boldsymbol{n}$ is positive towards the direction of the reactant mixture, such that,
\begin{equation} \label{eq:FFPT}
	S_d~\boldsymbol{n}= \frac{d \boldsymbol{x}_F}{dt} - \boldsymbol{u}
\end{equation}
If isotherms are used to represent flame surfaces, i.e., layers within the flame, then $S_d$ also denoted by $S_{d,T}$ is determined using the energy conservation equation \cite{poinsot2005}.

\begin{equation}
\begin{split}
    S_{d,T} & = \frac{1}{\rho C_p |\nabla T|}\left[ \nabla\cdot(\lambda^\prime\nabla T) + \rho \nabla T \cdot \sum_{k}^{}  (\mathcal{D}_k C_{p,k} \nabla Y_k) - \sum_{k}^{} h_k \dot{\omega}_k\right] \\
    \end{split}
    \label{eq: Sd_eq_DNS_T}
\end{equation}

\noindent where $\lambda^\prime$ is the thermal conductivity of the mixture at that isotherm; $\mathcal{D}_k$, $h_k$ and $\dot{\omega}_k$ are the mass diffusivity, enthalpy, and net production rate of the $k$th species at that isotherm, respectively. $C_{p,k}$ and $C_{p}$ denote the constant-pressure specific heat for the $k$th species and the bulk mixture using the mixture-averaged formula, respectively.

Within the finite flame structure, the variation of $S_d$ arising from density effects can be minimized by using a density-weighted flame displacement speed definition, $\widetilde{S_d} =\rho_0 S_d/\rho_u$, where $\rho_0$ is the density at the isotherm of interest and $\rho_u$ is the density of fresh reactants. Such a density-weighted flame displacement speed $\widetilde{S_d}$ is interpreted as the local mass flux of reactants into an isotherm per unit density of the fresh reactants. 

The flame displacement speed is a key input to the reduced-order models as in G-equation~\cite{williams1985} or flame surface density models~\cite{trouve1994} in predicting the flame propagation and subsequently the overall burning rates. The local $S_d$ and $\widetilde{S_d}$ are known to be affected by external fluid dynamic effects such as curvature, $\kappa$ \cite{echekki1996, echekki1999, im2000effects, chen1998correlation, cifuentes2018, chakraborty2005_1, uranakara2016, uranakara2017}, strain rates \cite{echekki1996, chakraborty2005_1}, and differential diffusion effects represented by the Lewis number, $Le$, \cite{haworth1992, rutland1993, chakraborty2005_2, alqallaf2019}.  
The correlations between the displacement speed and the local stretch have been extensively studied theoretically~\cite{pelce1982,matalon1982,matalon1983,candel1990,law2000,liang2017} and experimentally \cite{wu1985, egolfopoulos1989, law2000}. 
The relative strength of turbulence is commonly characterized by the turbulent Karlovitz number, $Ka$, defined as the ratio of the characteristic flame time to the Kolmogorov eddy turnover time. For low and moderate level of turbulent fluctuations ($Ka < 10)$),  the deviations of $\widetilde{S_d}$ from $S_L$ can be captured by defining two separate Markstein lengths as~\cite{giannakopoulos2015_1,giannakopoulos2015_2,dave2020}: 
 
\begin{equation} \label{eq: Sd_eq_3}
    \widetilde{S_d}(\theta^*) = S_L - \mathcal{L}_\mathbb{K}(\theta^*) 
    \mathbb{K} - \mathcal{L}_\kappa(\theta^*)S_L\kappa
\end{equation}

\noindent where $\mathbb{K}=a_T + S_d\kappa$ is the total stretch rate, $a_T$ is the tangential strain rate and $\kappa = \nabla\cdot\boldsymbol{n}$ is the curvature, and the non-dimensional temperature, $\theta^* = T_0/T_u$ with $T_0$ and $T_u$ being the temperature of the isotherm of interest and the unburned mixture, respectively. The Markstein length based on tangential strain rate ($\mathcal{L}_{\mathbb{K}}$) and curvature ($\mathcal{L}_\kappa$) are calculated from theoretical expressions, in the linear, weak stretch regime \cite{giannakopoulos2015_2}. 

At higher levels of turbulence, however, the correlations based on the weakly stretched flame theory reveal large errors \cite{im2016direct}. In a DNS study of moderate Karlovitz numbers ($Ka$), ~\citet{lee2022_2} found that the correlations are strong at the extreme points on the flame surface with mostly negative curvatures, but not so at locations with  positive curvatures. In another study, ~\citet{suillaud2022} proposed an improved correlation by using two effective turbulent Markstein lengths, which depend on both Lewis numbers ($Le$) and $Ka$, unlike the laminar ones. 

In general, the discrepancies from the weak stretch theory are mainly attributed to the large curvatures and the associated complex flame topology. \cite{ dunstan2013,griffiths2015,trivedi2019topology,brouzet2019}. For example, tunnel  formation (large positive curvature) and tunnel closure (large negative curvature)  were found to be a major contributor to the flame area loss for the reactant or product pockets, respectively \cite{ dunstan2013,griffiths2015,trivedi2019topology}. A more complex combination of different canonical topologies during the flame annihilation have also been observed \cite{brouzet2019}. The annihilation events namely island burnout, pinch-off and tunnel formation were found to be accompanied by large $S_d$ contributed majorly by the reaction term followed by the curvature term at large negative curvature by \citet{haghiri2020}. However, at low curvatures the enhanced $S_d$ was dominated by the reaction term. Apart from the complex flame annihilation events planar flame-flame interaction have also been studied in the past \cite{wichman1997,lu2003}. \citet{wichman1997} examined a model for interaction between two identical premixed planar laminar flames analytically using the governing equations and employed numerical solution of their model to confirm the results. Later, a similarity solution for planar flame-flame interaction for unity $Le$ was proposed by \citet{lu2003} under the constant density assumption. 


 Dave and Chaudhuri \cite{dave2020} conducted Lagrangian backward \cite{dave2018} and forward  \cite{chaudhuri2015} flame particle tracking analyses and found that large deviations $\widetilde{S_d} \! \gg \! S_L$ emerge from flame-flame interactions as they approach the annihilation events. To describe the phenomena, a theoretical analysis was conducted for interacting preheat zones in an imploding cylindrical premixed flame, yielding a correlation \cite{yuvraj2022local} 
\begin{equation}\label{eq: Sd_eq_II}
    \frac{\widetilde{S_{d}}}{S_L} = 1 -\frac{2 \widetilde{\alpha_0} \kappa}{S_L}
\end{equation} 
where $\widetilde{\alpha_0}=\rho_0 \alpha_0/\rho_u$ is the density-weighted thermal diffusivity at the isotherm of interest. Eq.~(\ref{eq: Sd_eq_II}) was found to yield a good correlation in moderate to intensely turbulent atmospheric pressure flames  \cite{yuvraj2022local} with $Le$ not too far from unity. Incidentally, the result is qualitatively consistent with that by Peters \cite{Peters2000}, $\widetilde{S_d} = \rho_0 S_L/\rho_u -\widetilde{\alpha_0} \kappa$, derived for both corrugated flamelet and thin reaction zone regimes. 

The simple linear correlation in Eq.~(\ref{eq: Sd_eq_II}) was found to properly describe the flame behavior at large negative curvature involving flame-flame interaction. Its applicability was further validated at high $Ka$ turbulent flame conditions \cite{yuvraj2022local,chaudhuri2022turbulent}. However,  for such a linear model to be adopted as a general correlation, an additional rigorous theoretical analysis is needed to determine its slope and intercept. Eq.~(\ref{eq: Sd_eq_II}) was derived based on a simplified model of a steady, inwardly propagating, cylindrical premixed flame under the assumption of constant density \cite{dave2020}. Moreover, the velocity of the unburned gas mixture was neglected, leading to zero convection for a constant-density mixture. A more comprehensive analysis to account for additional realism is warranted.

Therefore, the main objective of this study is to develop a comprehensive model for the flame speed correlations during the flame-flame interaction. A one-dimensional cylindrical interacting laminar flame is adopted as the model configuration similar to \citet{dave2020}, while incorporating additional physical aspects. For example, for a large negative curvature, the flame surfaces need not move at the same speed and hence the analysis becomes transient. Using DNS data, the choice of an imploding, cylindrical flame configuration to describe flame-flame interaction at large negative curvature will be justified. A finite flame thickness with density variation across the flame is also taken into account. Variable density renders the radial fluid flow velocity to be non-zero, and requires the inclusion of convection in the analysis of the preheat zone interaction. A more comprehensive and generalized flame speed correlations are then be derived and validated in detailed 1D simulations.

Subsequently, the derived correlations are validated in
a dataset of three-dimensional direct numerical simulations (DNS) over a range of Reynolds numbers ($Re_t$) and Karlovitz numbers ($Ka$) at atmospheric and elevated pressure. First, the statistical distribution of the principal curvatures on the flame surface is analyzed in order to justify that the adopted cylindrical model appropriately represent the realistic 3D turbulent flames. Following the insights from the local flame structure analysis, the local flame states associated with large negative curvatures in turbulence are represented as transient, highly curved, cylindrical, interacting, and laminar flame structures. A detailed comparative analysis of the results from the 3D DNS, 1D simulations and modeling is then performed. 

\section{1D simulations and theoretical model}\label{SS: derivation}
In this section, we present the detailed theoretical analysis along with the results from the 1D cases that will be crucial in explaining the dynamics of flame-flame interaction while carrying out the analysis. 
 
 \begin{table}[h!] \small
\footnotesize
\begin{center}
\begin{tabular}{ | p{4cm} | p{1.8cm} | p{1.8cm} | p{1.8cm} |} 
\hline
Parameters & P1C & P3C & P7C \\
\hline
\hline
$P$ [atm]& 1 & 3 & 7 \\
Domain dimensions [cm] & 0.800 & 0.600 & 0.400 \\
Grid points & 4000 & 6000 & 8000 \\
$S_L$ [cm/s] & 133.456 & 102.158 & 72.858 \\
$\delta_L [cm]$ & 3.485E-02 & 1.021E-02 & 4.599E-03 \\
$\delta_L/\Delta x$ & 174.250 & 102.100 & 91.980 \\
$\delta t$ [$\mu$s] & 1.000E-03 & 1.000E-03 & 1.000E-04\\
\hline
\end{tabular}
\caption{\label{tab:2}Details of the parameters for the 1D cases for the imploding flame studied in this paper. For all the cases, $T_u$ = 300 K, $\phi$ = 0.7, $\delta_L=(T_b^\circ - T_u)/|\nabla T|_{max}$.}
\end{center}
\end{table}

\begin{figure}[h!]
\centering\includegraphics[trim=2cm 14cm 1cm 2cm,clip,width=1.0\textwidth]{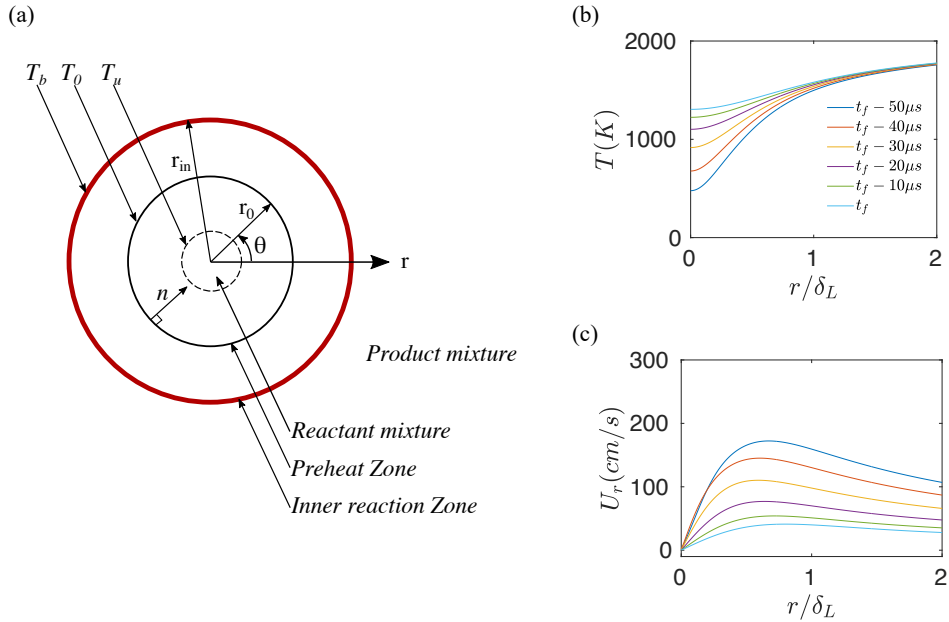}
\caption{(a) Schematic of the unsteady, imploding cylindrical laminar flame. (b) Radial temperature and (c) velocity profiles for the unsteady, imploding cylindrical laminar flame after every $10\mu s$ obtained from the 1D cylindrical laminar flame (P1C). $t_f$ is time instant at which $c_0 =0.6$ isotherm undergoes annihilation. The abscissa is normalized by the corresponding laminar flame thickness, $\delta_L$.}
\label{fig:1D_schematic}
\end{figure}


Three one-dimensional cases, namely, P1C, P3C and P7C, for inwardly propagating H$_2$-air cylindrical flame presented in Table~\ref{tab:2}, were simulated using the Pencil code \cite{thepencilcode}. The conservation equations of mass, momentum, energy and species are solved in a fully compressible formulation with NSCBC in the inflow-outflow direction. The Pencil code offers simulation in a cylindrical coordinate system as its major advantage, making it suitable for carrying out DNS of 1D cylindrical laminar premixed flame. The spatial discretization in the governing equations of mass, momentum and energy follows a sixth-order accurate central difference scheme. On the other hand, the fifth-order accurate upwind scheme is employed for the convection terms to remove any spurious oscillations in the flow field variables. A low-storage, third-order accurate Runge-Kutta RK3-2N scheme is used for marching the solution in time. A detailed chemical mechanism proposed by \citet{li2004} comprising of 9 species and 21 reactions is used to model the H$_2$-air chemistry.
Here the problem is configured such that the flow variables are invariant along the azimuthal direction rendering the flame propagation to be axisymmetric  and the problem one-dimensional. The flame is initially generated far from the center of the domain and allowed to develop in space while propagating towards the center. Once the flame (preheat isotherm) is approximately at a distance of $3\delta_L$ ($\delta_L$ being the laminar flame thickness), the solutions are saved at an interval of $0.2\mu s$.

Following \citet{dave2020}, Figure~\ref{fig:1D_schematic}a presents the schematic of an imploding, interacting isotherm (inner solid black circle) in the preheat zone far from the inner reaction zone (outer solid red circle), within a premixed flame. The unburned reactant mixture is enclosed within the innermost circle with the dotted curve. The suitability of cylindrical imploding flame in representing flame-flame interaction will be discussed later in section~\ref{S:4.1}.
In this study we use isotherms to represent layers within the flame and hence evaluate $S_d$ from the DNS using Eq.~(\ref{eq: Sd_eq_DNS_T}). We define the temperature-based progress variable $c$ as $c=(T-T_u)/(T_b^\circ - T_u)$. Figure~\ref{fig:1D_schematic}b and c present the temperature and velocity profiles obtained from the 1D simulations of imploding, cylindrical, laminar flame (P1C) captured with an increment of $10 \mu s$ during the interaction phase until the eventual annihilation of iso-scalar $c_0=0.06$. The 1D simulations reveal the transient nature of the thermal flame structure during the interaction phenomenon (Fig.~\ref{fig:1D_schematic}b). This emphasizes the fact that the physics of the flame-flame interaction is beyond the scope of the previously proposed linear models \cite{matalon1982,matalon1983,giannakopoulos2015_1,giannakopoulos2015_2} since they are based on the flame structure in a steady state. The temperature profile shows a decrease in $|\nabla T|$ with the progress in interaction, consistent with the local one-dimensional thermal structures obtained by \citet{dave2020} and \citet{yuvraj2022local} for moderately turbulent flames during flame-flame interaction.  Based on the cylindrical flame configuration, a simple analytical expression for the flame displacement speed is derived as follows.

The conservation equations for mass, energy, and species in cylindrical coordinates are written as:
\begin{equation}\label{eq: continuity}
    \frac{\partial \rho}{\partial t}+\frac{1}{r}\frac{\partial r \rho U_r}{\partial r} = 0
\end{equation}
%

\begin{equation}\label{eq: energy_conservation_3}
  \rho C_{p} \left( \frac{\partial T}{\partial t} + U_r\frac{\partial T}{\partial r} \right)=\frac{1}{r}\frac\partial{\partial r}\left(\lambda^{\prime} r \frac{\partial T}{\partial r} \right) + qw
\end{equation}
\begin{equation}\label{eq: mass_conservation_3}
  \rho \left( \frac{\partial Y}{\partial t} + U_r \frac{\partial Y}{\partial r} \right)=\frac{1}{r}\frac{\partial}{\partial r}\left(\rho \mathcal{D} r \frac{\partial Y}{\partial r} \right) - w
\end{equation}
where $q$ and $w$ denote the heat release and reaction rate; $C_p$, $U_r$, and $\lambda^{\prime}$ are the heat capacity, the radial flow velocity, and the thermal conductivity, respectively. From Eq.~(\ref{eq: continuity}) it is clear that accounting for the variation in $\rho$ leads to variable $U_r$. Hence, the effect of variable density is embedded in the form of non-zero convection term in Eq.~(\ref{eq: energy_conservation_3}) and (\ref{eq: mass_conservation_3}).

The energy equation is first solved in the preheat zone accounting for the motion of the isotherm lying in the reaction zone. In the present analysis, we assume the thermal diffusivity, $\alpha_0 = \lambda^\prime/(\rho C_p)$, to be constant. The temperature and the radius of the isotherm of interest are set as $T_{0}$ and $r_0(t)$, respectively. Capturing the motion of the isotherm surface depends on the location of the inner reaction layer. Therefore, it is necessary to define Eq.~(\ref{eq: energy_conservation_3}) in a new coordinate system to consider the reaction zone's motion.

Introducing the non-dimensional temperature, $\theta= (T_{0}-T)/(T_{0}-T_{u})$ where $T_{u}$ is the temperature of unburned gas mixture, and applying the coordinate transformation to two independent variables $\xi = r/r_0(\tau)$ and $\tau = t$, Eq.~(\ref{eq: energy_conservation_3}) is  expressed as given by Eq.~(\ref{eq: energy_conservation_stretch}), where $\tau$ is renamed to $t$ for simplicity:

\begin{equation}\label{eq: energy_conservation_stretch}
\begin{aligned}
  \frac{\partial \theta}{\partial t}- \frac{\xi}{r_0} \left( \frac{dr_0}{dt} - \frac{U_r}{\xi} \right) \frac{\partial \theta}{\partial \xi}=\frac{\alpha_0}{r_0^{2}} \left[\frac{1}{\xi} \frac{\partial \theta}{\partial \xi} + \frac{\partial^{2} \theta}{\partial \xi^{2}} \right]-\frac{qw}{\rho C_p (T_0-T_u)}
\end{aligned}
\end{equation}

Furthermore, as seen from Fig.~\ref{fig:1D_schematic}c, $U_r=0$ at $r=0$ and $U_r=U_{r,0}$ at $r=r_0$. Moreover, $U_r$ can be approximated as $\xi U_{r,0}$ in the domain $0\leq \xi \leq1$ as a leading approximation (see Fig.~\ref{fig:1D_schematic}). 
Considering that ${dr_0}/{dt} = U_{r,0} - S_{d}$ \cite{Peters2000}, Eq.~(\ref{eq: energy_conservation_stretch}) becomes,

\begin{equation}\label{eq: energy_conservation_stretch_4}
\begin{aligned}
  \frac{\partial \theta}{\partial t} + S_{d}\frac{\xi}{r_0} \frac{\partial \theta}{\partial \xi}=\frac{\alpha_0}{r_0^{2}} \left[\frac{1}{\xi} \frac{\partial \theta}{\partial \xi} + \frac{\partial^{2} \theta}{\partial \xi^{2}} \right]-\frac{qw}{\rho C_p (T_0-T_u)}
\end{aligned}
\end{equation}

Multiplying Eq.~(\ref{eq: energy_conservation_stretch_4}) by $\xi$ and take an integral over [0,1] with respect to $\xi$, 
\begin{equation}\label{eq: energy_conservation_stretch_5}
\begin{aligned}
  \int_{0}^{1} \xi \theta_t d\xi + \frac{S_{d}}{r_0} \int_{0}^{1} \xi^2 \theta_\xi d\xi=
  \frac{\alpha_0}{r_0^{2}} \left[\int_{0}^{1} \theta_\xi d\xi + \right. \\ \left.\int_{0}^{1} \xi \theta_{\xi\xi} d\xi \right]  -\int_{0}^{1} \frac{qw}{\rho C_p (T_0-T_u)} \xi d\xi
\end{aligned}
\end{equation}
where subscripts $\xi$ and $t$ denote a short-hand notation for derivatives. The known boundary conditions are $\theta = 1$ at $\xi = 0$, and $\theta = 0$ at $\xi = 1$. The integral is further simplified by using an approximate temperature profile,  $\theta(\xi)=(\xi-1)\theta_\xi(1)$ at leading order, such that
\begin{equation}\label{eq: deriv_theta_estimation_xi}
  \theta_\xi(\xi)=\theta_\xi(1), \theta_t(\xi)=(\xi-1)\theta_{t\xi}(1)
\end{equation}
and further considering that the reaction term is zero in the range $0\leq \xi \leq1$,  Eq.~(\ref{eq: energy_conservation_stretch_5}) yields,
\begin{equation}\label{eq: energy_conservation_grad_xi}
  \frac{\theta_{t\xi}(1)}{6} - \frac{S_{d}}{r_0} \frac{\theta_{\xi}(1)}{3}=-\frac{\alpha_0}{r_0^{2}}\theta_{\xi}(1)
\end{equation}
which serves as a balance of various transport terms at the front edge of the reaction zone. 

\begin{figure}[h!]
\centering\includegraphics[trim=5.2cm 4.5cm 5.2cm 7cm,clip,width=1.0\textwidth]{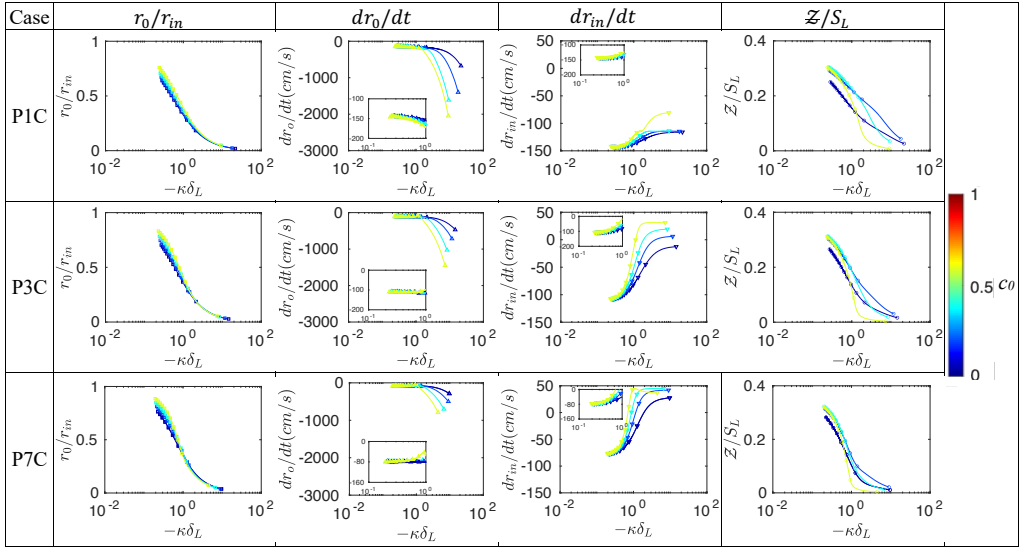}
\vspace{10 pt}
\caption{1D profiles of $r_0/r_{in}$, $dr_0/dt$, $dr_{in}/dt$ and $\mathcal{Z}/S_L$ with $-\kappa\delta_L$ for $c_0=0.05, 0.2, 0.4$ and $0.6$ based on the colorscale included alongside for all cases. The abscissa is in the log scale. $r_0$ and $r_{in}$ are the radius of isotherm of interest ($T_0$) and the inner reaction layer ($T_{in}$) respectively. Here we take $T_{in}=1800$ K.}
\label{fig:1D_ro_rin}
\end{figure}

We now consider the effect of the non-interacting, steadily propagating reaction zone. The upstream edge of the reaction zone is referred to as the \emph{inner layer} located at $r = r_{in}$, i.e., $\xi=\xi_{in}$ where $T = T_b^\circ$ at leading order, while $T_b^\circ$ is the adiabatic flame temperature in the unity $Le$ limit.
Similarly, $\theta$ at the inner layer becomes $\theta({\xi_{in}})= (T_{0}-T_{b}^\circ)/(T_{0}-T_{u})$ 
and assuming a linear temperature profile within $1 < \xi < \xi_{in}$, $\theta_{\xi}(1)\approx \theta(\xi_{in})/(\xi_{in}-1)$. Although $T_b^\circ$ and $\theta(\xi_{in})$ is constant in the model configuration, the inner layer moves steadily while the preheat zones interact. Thus, $r_{in}$ and $\xi_{in}$ remain functions of time. Taking time derivative of $\theta_{\xi}(1)$,
\begin{equation}\label{eq: deriv_theta_estimation_t_in_1}
 \theta_{t\xi}(1)\approx-\frac{\theta(\xi_{in})}{(\xi_{in}-1)^2} \frac{\partial \xi_{in}}{\partial t}
\end{equation}
where
\begin{equation}\label{eq: grad_xi_in}
 \frac{\partial \xi_{in}}{\partial t}= \frac{1}{r_0}\frac{dr_{in}}{dt}-\frac{\xi_{in}}{r_0}\frac{dr_0}{dt}
\end{equation}
by definition. Eq.~(\ref{eq: grad_xi_in}) is important for a cylindrical flame with a small radius and represents the interaction dynamics. Figure~\ref{fig:1D_ro_rin} shows the profiles of $r_0/r_{in}$, $dr_0/dt$, $dr_{in}/dt$ and $\mathcal{Z}/S_L$ (to be defined and discussed later) with $-\kappa\delta_L$ for the 1D cases P1C, P2C and P3C for $c_0 = 0.05, 0.2, 0.4$ and $0.6$ with the color scale representing the corresponding value of $c_0$. For an inwardly propagating flame of a large radius or small curvature, the $r_{in} \approx r_0 >> \delta_L$, i.e., $r_{0}/r_{in}\approx 1$. At the same time, both $dr_{0}/dt$ and $dr_{in}/dt$ are also approximately equal and constant. The closeup views in Fig.~\ref{fig:1D_ro_rin} (second and third column) show that $dr_{0}/dt$ and $dr_{in}/dt$ are approximately equal to -150, -100 and -80 for the cases P1C, P3C and P7C respectively in the non-interacting limit. Hence, we can assume $\xi_{in} dr_0/dt$ $\approx$ $dr_{in}/dt$ based on  the 1D results for fairly small values of $-\kappa\delta_L$ (approximately less than or equal to 0.1 in Fig.~\ref{fig:1D_ro_rin}). Therefore, from Eq.~(\ref{eq: grad_xi_in}) $\partial \xi_{in}/\partial t \approx 0$ resulting in a steady-state problem. 

However, this is not the case when the flame surfaces undergo interaction. When the isotherm of interest ($T_0$) lying in the preheat zone interacts, its radius of curvature $r_0$ is appreciably small in comparison to the radius of the reaction layer $r_{in}$ as shown for the 1D case in Fig.~\ref{fig:1D_ro_rin} (first column). With the progression in interaction and decrease in $\kappa$, $r_0/r_{in} \rightarrow 0$. During the initial stage of interaction, the magnitude of $dr_0/dt$ increases (second column in Fig.~\ref{fig:1D_ro_rin}), whereas that of $dr_{in}/dt$ decreases (third column in Fig.~\ref{fig:1D_ro_rin}). As the interaction proceeds ($\kappa\delta_L \! \ll \! -1$ and $r_0/r_{in} \rightarrow 0$), the inner layer, $dr_{in}/dt$, levels off to a constant value. On the other hand, $dr_0/dt$ shows a sharp increase in magnitude. In reality, however, $dr_{in}/dt$ is not constant but changes negligibly throughout the duration of the interaction until the isotherm of interest eventually annihilates. This also confirms that the interaction phase is very short-lived. 
Implementing the insights gained from 1D cases during the interaction phase to our theory in Eq.~(\ref{eq: grad_xi_in}), $\partial\xi_{in}/\partial t \neq 0$ justifying the transient nature of the interacting flames.

To eliminate $\partial\xi_{in}/\partial t$ using Eq.~(\ref{eq: deriv_theta_estimation_t_in_1}) and Eq.~(\ref{eq: grad_xi_in}), we substitute $\theta_{t\xi}$ into Eq.~(\ref{eq: energy_conservation_grad_xi}),

\begin{equation}\label{eq: energy_conservation_stretch_7}
 \frac{r_0}{r_{in}-r_0}\frac{dr_{in}}{dt}-\frac{r_{in}}{r_{in}-r_0}\frac{dr_0}{dt} + 2S_{d}=\frac{6\alpha_0}{r_0}
\end{equation}

Since, $r_0 \! \ll \! r_{in}$ in the interaction phase we obtain a simplified expression for $S_{d}$ as, 

\begin{equation}\label{eq: energy_conservation_stretch_8}
 \frac{r_0}{r_{in}}\frac{dr_{in}}{dt}-\frac{dr_0}{dt} + 2S_{d}=\frac{6\alpha_0}{r_0}
\end{equation}
 
Again, substituting $dr_0/dt$ using $\frac{dr_0}{dt} = U_{r,0} - S_{d}$ in Eq.~(\ref{eq: energy_conservation_stretch_8}), the equation for $S_d$ is obtained as,

\begin{equation}\label{eq: drf/dt_interaction_negative_7}
  S_{d}= \frac{2\alpha_0}{r_0} + 
  \frac{1}{3} \left[ U_{r,0} - \frac{r_0}{r_{in}}\frac{dr_{in}}{dt} \right]
\end{equation}

\noindent where $S_{d}$ is defined as the displacement speed of the isotherm $T = T_0$, whereas the $-dr_{in}/dt$ is the speed of the inner layer, $\kappa=-1/r_0$, and $U_{r,0}$ is the radial gas velocity at the isotherm $T = T_0$. Performing density weighing of Eq.~(\ref{eq: drf/dt_interaction_negative_7}) we have

\begin{equation}\label{eq: drf/dt_interaction_negative_8}
  \widetilde{S_{d}}= -2\widetilde{\alpha_0}\kappa + 
  \frac{1\rho_0}{3\rho_u} \left[ U_{r,0} - \frac{r_0}{r_{in}}\frac{dr_{in}}{dt} \right]
\end{equation}


\noindent where $\widetilde{\alpha_0}$ and $\widetilde{S_d}/S_L$ denote the density-weighted thermal diffusivity $\widetilde{\alpha_0}=\rho_0 \alpha_0/\rho_u$ and flame displacement speed $\widetilde{S_d}=\rho_0 S_d/\rho_u$ respectively. We denote the second term on the right-hand side of Eq.~(\ref{eq: drf/dt_interaction_negative_8}) as

\begin{equation}\label{eq: intercept}
  \mathcal{Z}= \frac{1\rho_0}{3\rho_u} \left[ U_{r,0} - \frac{r_0}{r_{in}}\frac{dr_{in}}{dt} \right]
\end{equation}


In Eq.~(\ref{eq: drf/dt_interaction_negative_8}), we get a finite intercept, $\mathcal{Z}$ when accounting for the gas velocity and the motion of the inner layer. However, in the interacting configuration at very large negative $\kappa$, $\mathcal{Z}\rightarrow0$ as shown in Fig.~\ref{fig:1D_ro_rin} (fourth column). Thus, $\widetilde{S_d}$ is independent of $U_{r,0}$ at large negative $\kappa$ even after the inclusion of density variation in the current analysis. Consequently, Eq.~(\ref{eq: drf/dt_interaction_negative_8}) reduces to $\widetilde{S_d} = -2\widetilde{\alpha_0}\kappa$ as obtained by Dave et al. \cite{dave2020}. However, in the non-interacting limit as $\kappa\rightarrow 0 $, $\widetilde{S_{d}}$ approaches some finite intercept $\widetilde{S_{d,0}}$. 
This yields:

\begin{equation}\label{eq: final_density-weighted_Sd_kappa_S_L*}
  \widetilde{S_{d}}=\widetilde{S_{d,0}} -2 \widetilde{\alpha_0} \kappa
\end{equation}

As the simplest approximation, one can consider $\widetilde{S_{d,0}} \rightarrow S_L$ to the leading order for flames with $Le$ close to unity. Hence,

\begin{equation}\label{eq: final_density-weighted_Sd_kappa}
  \widetilde{S_{d}}=S_L -2 \widetilde{\alpha_0} \kappa
\end{equation}


Although the intercept $\widetilde{S_{d,0}}$ may deviate from $S_L$ due to non-unity $Le$ and turbulence effects (discussed later in section~\ref{S:4.3}), the present derivation is more generalized without many restrictive assumptions and the same composite result as Eq.~(\ref{eq: final_density-weighted_Sd_kappa}) proposed by Yuvraj et al. \cite{yuvraj2022local} is recovered. Therefore, a model applicable in both the interacting and non-interacting configurations for $\kappa\leq0$ is obtained as Eq.~(\ref{eq: final_density-weighted_Sd_kappa}).

\section{DNS Cases}
\label{S:2}
The theoretical model and correlations derived in the previous section are now assessed in the context of realistic turbulent flames by using the DNS dataset. The three-dimensional DNS cases investigated in the current work are listed in Table~\ref{tab:1}. Atmospheric pressure cases were generated for the previous work by \citet{song2020dns} and \citet{yuvraj2022local}, whereas the elevated pressure cases are from Song et al. \cite{song2022diffusive}. The simulations were performed with lean H$_2$-air premixed flame propagating into forced turbulence in a cuboidal domain with turbulence by varying the turbulence parameters. Two out of the four data sets under consideration, i.e., F1, and F2, are at atmospheric pressure. The remaining data sets, P3 and P7, are at an elevated pressure of 3 and 7 atm, respectively.

\begin{table}[h!]
\footnotesize
 \begin{center}
 \begin{tabular}{ | p{4cm} | p{1.6cm} | p{1.6cm} | p{1.6cm} | p{1.6cm} |} 
 \hline
 Parameters & F1 & F2 & P3 & P7 \\ 
  \hline
  \hline
  $P$ [atm] & 1 & 1 & 3 & 7 \\
  $Le$ & 0.408 & 0.408 & 0.408 & 0.408 \\
$Le_{eff}$~\cite{matalon_2003} & 0.758 & 0.758 & 0.698 & 0.656 \\
 Domain dimensions [cm] & $1.998$ & 0.427& 0.647 & 0.656 \\
  &$\times1.000$ & $\times0.146$ & $\times0.215$ & $\times0.218$\\
  &$\times1.000$ & $\times0.146$ & $\times0.215$ & $\times0.218$\\
 Grid points & $1000$ & $1645$ & $648$ & $1440$ \\
  & $\times500$ & $\times560$ & $\times216$ & $\times480$ \\
  & $\times500$ & $\times560$ & $\times216$ & $\times480$ \\
 Integral length scale, $l_o$ [cm] & 0.200 & 0.029 & 0.043 & 0.043 \\
 Root mean square velocity, $u_{rms}$ [cm/s] & 678.096 & 4746.669 & 317.644   & 225.699 \\
 Kolmogorov length scale, $\eta$ [$\mu$m] & 14.915 & 2.141 & 7.921 & 5.421 \\
 Karlovitz no., $Ka$ &  23.189 &  1125.660 & 10.243 & 5.888 \\
 Reynolds no., $Re_t$& 686.350 & 699.530 & 204.225 & 338.590 \\
 Damk\" ohler no., $Da$ & 1.130 & 0.023 & 1.395 & 3.125 \\
 $S_L$ [cm/s] & 135.619 & 135.619 & 105.882 & 75.233 \\
 $\delta_L$ [cm] & 3.541E-02 & 3.541E-02 & 1.022E-02 & 4.565E-03 \\
$\Delta x/\eta$ & 1.341 & 1.215 & 1.263 & 0.841 \\
$\delta_L/\Delta x$ & 17.705 & 136.192 & 10.224 & 10.010 \\
$\delta t$ [$\mu$s] & 1.000E-02 & 1.000E-03 & 2.500E-03 & 1.250E-03\\
 \hline
 \end{tabular}
 \caption{\label{tab:1}Details of the parameters for the four 3D DNS cases studied in this paper. For all the cases, $T_u$ = 300 K, $\phi$ = 0.7, $\delta_L=(T_b^\circ - T_u)/|\nabla T|_{max}$}
 \end{center}
 \end{table}

\begin{figure}[h!]
\centering\includegraphics[trim=2.5cm 9cm 2cm 8.1cm,clip,width=1.0\textwidth]{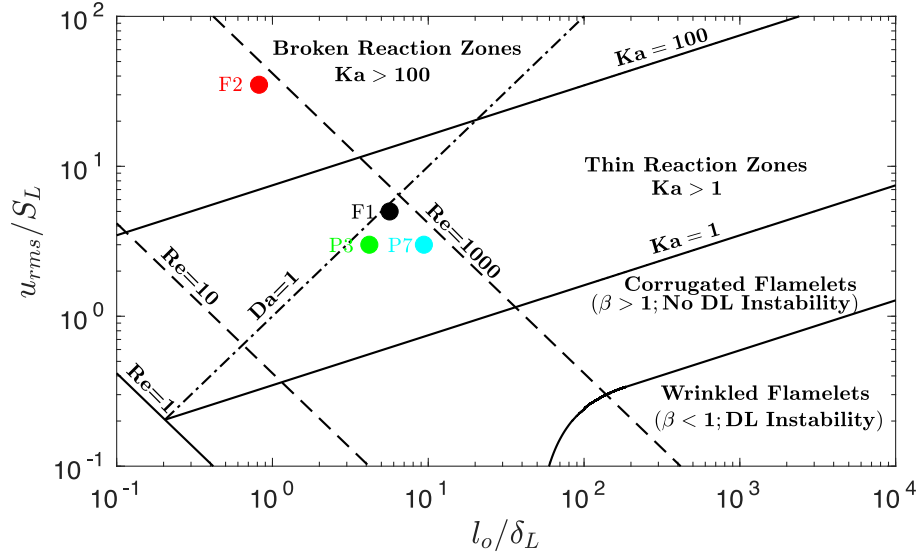}
\caption{Regime diagram specifying the cases: F1, F2, P3, P7.}
\label{fig:regime_diagram}
\end{figure}

Figure~\ref{fig:regime_diagram} shows  the modified Borghi-Peters turbulent regime diagram~\cite{chaudhuri2011}, indicating the simulation cases based on the integral length scales $l_0$ and turbulence intensities $u_{rms}$ at input conditions, resulting in a variety of turbulent Reynolds ($Re_t$) and Karlovitz numbers ($Ka$). $\delta_L$ is the corresponding thermal thickness of the standard laminar flames defined as $\delta_L=(T_b^\circ - T_u)/|dT/dx|_{max}$ with $T_b^\circ$ being the adiabatic flame temperature. Cases F1, P3 and P7 lie in the thin reaction zone, whereas F2 is located in the broken reaction zone regime. 

KAUST Adaptive Reacting Flow Solver (KARFS)~\cite{perez2018direct} was used to solve the conservation equations of mass, momentum, energy and species in a fully compressible formulation. For spatial discretization, an eighth-order central-difference scheme is used whereas a fourth-order Runge-Kutta scheme is used for time integration. Alongside, spurious fluctuations are removed using a tenth-order filter. At the outlet, the non-reflecting Navier-Stokes characteristic boundary conditions (NSCBC)~\cite{yoo2005characteristic,yoo2007characteristic} are applied whereas periodic boundary conditions exist in the transverse directions. A detailed mechanism comprising of 9 species and 23 reactions from \cite{burke2012comprehensive} is used.


\section{Results and Discussion}
\label{S:4}

\subsection{Principal curvatures and principal planes}
\label{S:4.1}
\begin{figure}[h!]
\centering\includegraphics[trim=2.2cm 11.5cm 2.2cm 2.5cm,clip,width=1.0\textwidth]{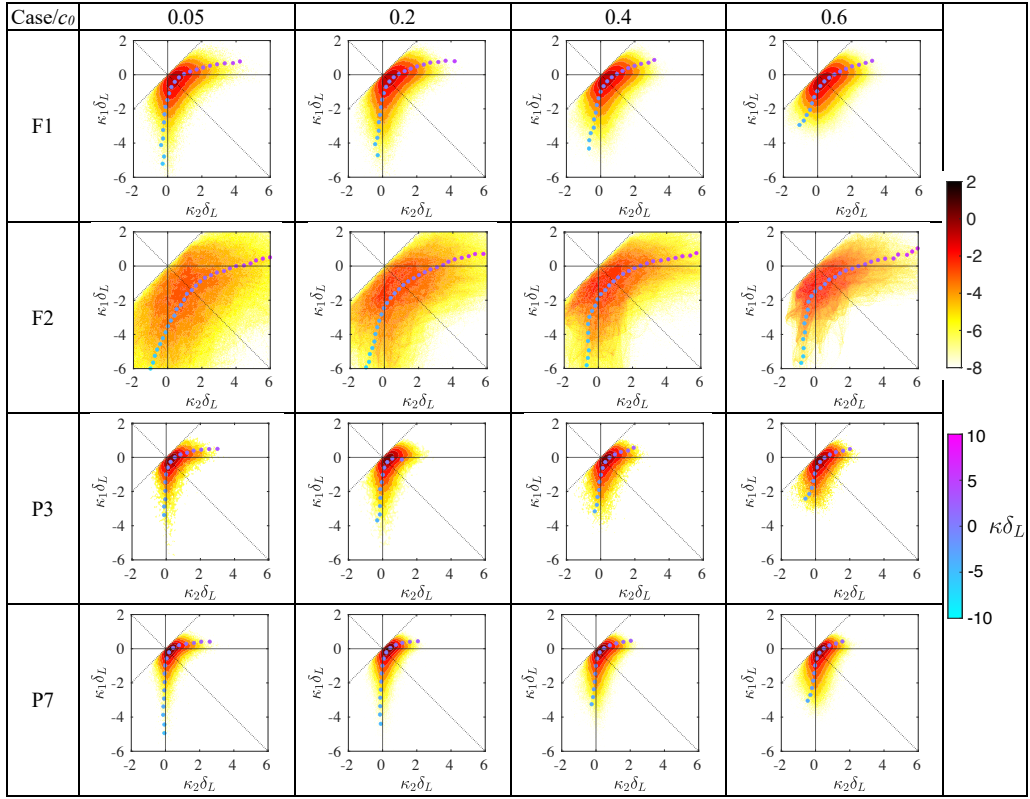}
\caption{Joint probability density function (JPDF) of normalized principal curvatures $\kappa_1 \delta_L$ and $\kappa_2 \delta_L $ ($\kappa_1 \delta_L$ $<$ $\kappa_2 \delta_L $) for $c_0$ = 0.05, 0.2, 0.4 and 0.6 over four time instants. Colorscale (top) represents the natural logarithm of JPDF magnitudes. The filled circles represent the set of points ($\langle\kappa_2\delta_L|_{\kappa\delta_L}\rangle,\langle\kappa_1\delta_L|_{\kappa\delta_L}\rangle$) given $\langle\kappa_2\delta_L|_{\kappa\delta_L}\rangle+\langle\kappa_1\delta_L|_{\kappa\delta_L}\rangle=\kappa\delta_L$; the markers are colored based on the values of $\kappa\delta_L$ (bottom colorscale).}
\label{fig:JPDF_kappa1_kappa2}
\end{figure}

At any given point on an iso-scalar surface within a flame, two unique, mutually perpendicular planes containing the vectors ($\boldsymbol{n}$, $\boldsymbol{e_1}$) and ($\boldsymbol{n}$, $\boldsymbol{e_2}$) can be formed such that the local curvatures of the surface in these planes are minimum and maximum, respectively. The curvature values thus obtained are called the principal curvatures ($\kappa_1<\kappa_2$) and the planes are referred to as the principal planes. In the present study, the curvature $\kappa$ at a given point on the flame surface is the sum of principal curvatures, 
 \begin{equation} \label{eq:principal_curvature}
     \kappa = \nabla\cdot\boldsymbol{n} = \kappa_1 + \kappa_2	
 \end{equation}
where $\boldsymbol{n}$ is the unit normal vector and $\boldsymbol{e_1}$, $\boldsymbol{e_2}$ are two unit vectors perpendicular to each other on the surface, in the direction of the local maximum and minimum curvature, respectively. All three unit vectors $\boldsymbol{n}$, $\boldsymbol{e_1}$ and $\boldsymbol{e_2}$ at a given point are orthogonal by definition. 

Figure~\ref{fig:JPDF_kappa1_kappa2} presents the joint probability density function (JPDF) of non-dimensional principal curvatures, $\kappa_1\delta_L$ and $\kappa_2\delta_L$ for four different iso-$c$ surfaces. $c_0 = $ 0.05, 0.2, 0.4 and 0.6 are chosen for investigation for all cases. The color map shows the natural logarithm of the JPDF magnitudes. $\kappa_1$ is zero along the horizontal solid line and $\kappa_2$ is zero along the vertical solid line. $\kappa_1 = \kappa_2$ along the dotted line with $+1$ slope. $\kappa_1 = -\kappa_2$ along the dotted line with $-1$ slope. $\langle\kappa_1\delta_L|_{\kappa\delta_L}\rangle$ and $\langle\kappa_2\delta_L|_{\kappa\delta_L}\rangle$ are the conditional means of the respective non-dimensional principal curvatures given total non-dimensional curvature $\kappa\delta_L$. The set of points ($\langle\kappa_2\delta_L|_{\kappa\delta_L}\rangle$, $\langle\kappa_1\delta_L|_{\kappa\delta_L}\rangle$) for which $\langle\kappa_2\delta_L|_{\kappa\delta_L}\rangle+\langle\kappa_1\delta_L|_{\kappa\delta_L}\rangle=\kappa\delta_L$ are shown in filled circles on each of the JPDFs.
The colors of the circle denote the magnitude of normalized curvature, $\kappa\delta_L$. So, if a filled circle intersects the dotted line with slope = -1, it implies, statistically, the local surfaces are hyperboloid. If the filled circle intersects with one of the axes, the local surfaces are cylindrical. 

On all surfaces across all cases (atmospheric as well as elevated pressure), it is evident that an extreme magnitude of one principal curvature simultaneously is paired with a small (near zero) magnitude of the other.  More importantly, the extreme value of local curvature is contributed mostly by one of the principal curvatures. The high $Ka$ case F2 indeed shows a large scatter in $\kappa_1$ and $\kappa_2$ values but the overall nature of the distribution remains the same. The high probability of the presence of large negatively or positively curved near cylindrical flame surfaces is consistent with \citet{shim2011}. However, the likelihood of very large curvatures is reduced as $c_0$ increases or moves toward the product side of the flame. It is evident from Fig.~\ref{fig:JPDF_kappa1_kappa2} that, at least for $c_0\leq0.4$, the largest negative (positive) curvatures are primarily contributed by $\kappa_1$ ($\kappa_2$) and hence those local flame surfaces are mostly cylindrical rather than spherical. This justifies the choice of a cylindrical configuration for locally analyzing highly negatively curved turbulent flame structures.
It should also be noted that $\langle\kappa_1\delta_L|_{\kappa\delta_L=0}\rangle$ and $\langle\kappa_2\delta_L|_{\kappa\delta_L=0}\rangle$ are non-zero. This assures on the fact that, while one might expect the flame surfaces to be planar at $\kappa\delta_L=0$ on average, these surfaces are in reality hyperboloid, with $\langle\kappa_1\delta_L|_{\kappa\delta_L=0}\rangle+\langle\kappa_2\delta_L|_{\kappa\delta_L=0}\rangle=0$. Local collection of such saddle points with negative Gaussian curvature ($\kappa_1\kappa_2<0$) together form the local saddle surface i.e., the hyperbolic hyperboloid. Nevertheless, it can be stated that locally planar flame surfaces ($\kappa_1\delta_L=\kappa_2\delta_L =0$) do exist but are less prevalent.   

\begin{figure}[h!]
\centering\includegraphics[trim=0cm 11.5cm 0cm 6cm,clip,width=1.0\textwidth]{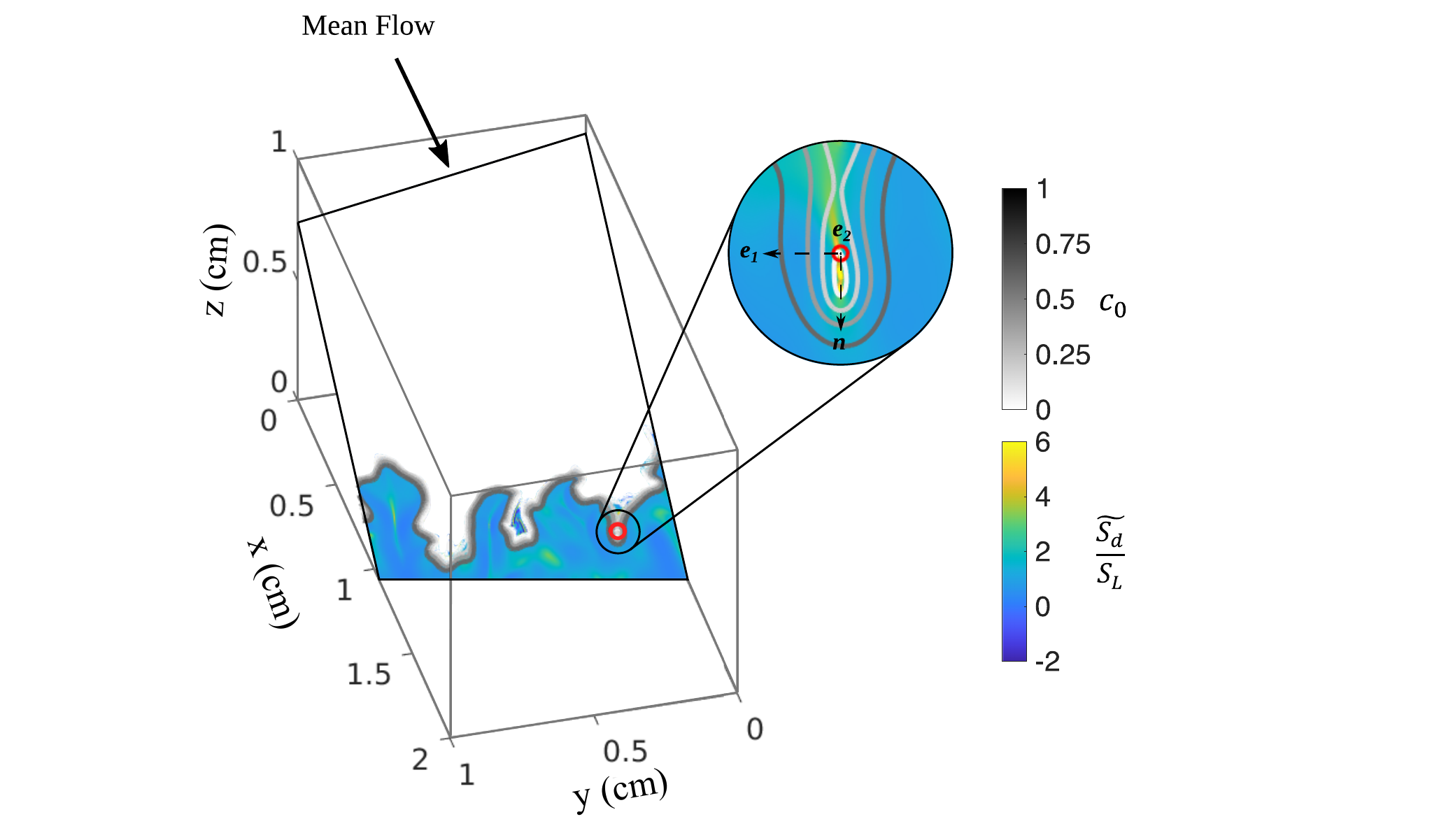}
\vspace{10 pt}
\caption{Schematic showing the principal plane with a minimum value of curvature at the point of interest (red circle) containing vectors $\boldsymbol{n}$ and $\boldsymbol{e_1}$. The vector $\boldsymbol{e_2}$ coincides with the normal vector of the plane. The colorbars represent the progress variable, $c_0$ (top) and the normalized density weighted flame displacement speed, $\widetilde{S_d}/S_L$ (bottom).}
\label{fig:schematic_principal_plane}
\end{figure}

\begin{figure}[h!]
\centering\includegraphics[trim=2.5cm 9.6cm 2.5cm 4cm,clip,width=1.0\textwidth]{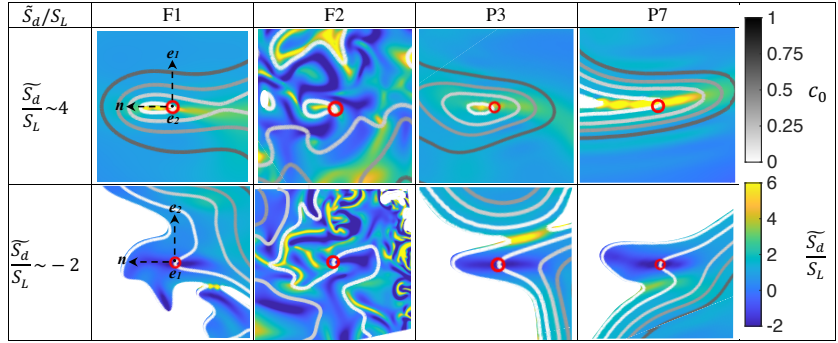}
\vspace{10 pt}
\caption{Instantaneous principal plane contours of normalized density weighted flame displacement speed, $\widetilde{S_d}/S_L$. The top row shows the principal plane (at the point marked with a red symbol) with a minimum value of curvature $\kappa_1 \delta_L$ containing the vectors $\boldsymbol{n}$ and $\boldsymbol{e_1}$ whereas the bottom row shows the principal plane with a maximum value of curvature ($\kappa_2 \delta_L$) containing the vectors $\boldsymbol{n}$ and $\boldsymbol{e_2}$. The solid lines show the two-dimensional isolines for the progress variable, $c_0 =$ 0.01, 0.05, 0.2, 0.4 and 0.6. The point with enhanced $\widetilde{S_d}/S_L$ ($\sim4$) due to large negative $\kappa_1 \delta_L$ and negative $\widetilde{S_d}/S_L$ ($\sim-2$) due to large positive $\kappa_1 \delta_L$ is denoted by red circle in the top and bottom row for each case. Each of the contours is enclosed in a square cutout of edge length $3\delta_L$. The colorbars represent $c_0$ (top) and $\widetilde{S_d}/S_L$ (bottom).}
\label{fig:2D_structures_Sd}
\end{figure}


Since the magnitude of large curvatures is often dominated by one of the principal curvatures, the corresponding local flame structures become quasi cylindrical, such that the associated, density-weighted, normalized flame speed $\widetilde{S_d}/S_L$ is visualized in the corresponding principal planes. 
The orientation of the principal plane depends on the direction of the local unit vectors $\boldsymbol{n}$, $\boldsymbol{e_1}$ and $\boldsymbol{e_2}$. The principal plane (with large negative $\kappa_1$) at one such point in the computational domain for Case F1 is shown in Fig.~\ref{fig:schematic_principal_plane} along with the color map indicating he magnitude of $\widetilde{S_d}/S_L$ and the progress variable $c_0$. The close-up view of the enhanced $\widetilde{S_d}$ at large negative $\kappa_1$ is also shown. 

Following the analysis, Fig.~\ref{fig:2D_structures_Sd} shows a number of samples of the two-dimensional flame structure on these principal planes, around a specific point of interest (POI) where the maximum curvature regions are marked by small red circles on the $c_0= 0.05$ iso-surfaces. The top and bottom rows, respectively, show the principal planes
containing ($\boldsymbol{n}$,  $\boldsymbol{e_1}$) and ($\boldsymbol{n}$, $\boldsymbol{e_2}$). For all the selected points of interest, $\widetilde{S_d}/S_L\sim4$ (top row) and $\widetilde{S_d}/S_L\sim-2$ (bottom row), the two-dimensional iso-$c_0$ lines ($c_0= 0.01,0.05,0.2,0.4$ and $0.6$) are superimposed on the principal plane. Since the iso-$c_0$ surface at most of these points is predominantly cylindrical, as discussed earlier, the point lying on $c_0=0.05$ is selected randomly from the filtered set. The top row clearly shows local cylindrical flame-flame interaction with a large negative $\kappa_1$, leading to a large positive $\widetilde{S_d}/S_L$ at and around the POI.  The bottom row shows the distortion of the preheat layers into positively curved structures with negative $\widetilde{S_d}/S_L$. The $\widetilde{S_d}/S_L$ varies significantly along the $c_0$ isocontour since $|\nabla c|$ is very small in this region. 
 
Figure~\ref{fig:2D_structures_Sd} shows that global flame-flame interaction occurs for cases F1, P3 and P7, whereas for case F2 with high $Ka$ turbulence, iso-surfaces are not parallel to each other, so that the flame-flame interaction occurs in a more localized manner, as discussed by Yuvraj et al. \cite{yuvraj2022local}. As discussed by \citet{gran1996negative}, the negative displacement speed results from a mismatch between diffusive and convective fluxes. Both large positive and negative curvature cases involve highly transient phenomena with large disruptions in the flame structure that deviates from the behavior of weakly stretched laminar flames. 

Thus, during these transient events, the configuration of an imploding cylindrical flame configuration (as presented before) is an useful (though not exhaustive) representation of local flame surfaces undergoing flame flame interaction at large negative curvatures.

\subsection{Joint Probability Density Functions of \texorpdfstring{$\widetilde{S_d}/S_L$}{} and \texorpdfstring{$\kappa\delta_L$}{}}

\begin{figure}[h!]
\centering\includegraphics[trim=2cm 11cm 3cm 3.3cm,clip,width=1.0\textwidth]{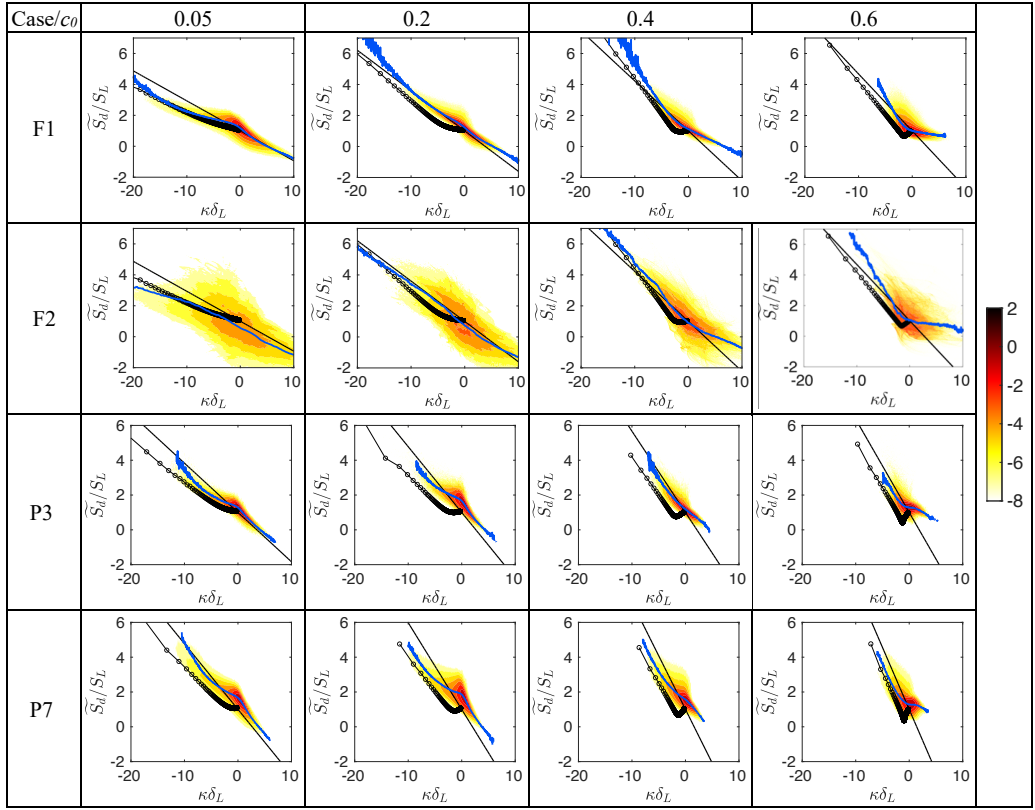}
\vspace{10 pt}
\caption{Joint probability density function (JPDF) of normalized density weighted flame displacement speed, $\widetilde{S_d}/S_L$ and normalized curvature $\kappa \delta_L$ for iso-scalar, $c_0 =$ 0.05, 0.2, 0.4 and 0.6 over four-time instants. Colorscale represents the natural logarithm of JPDF magnitudes. The solid black line represents the analytical results given by Eq.~(\ref{eq: final_density-weighted_Sd_kappa}). The 1D results are shown in a solid black curve with circular markers and the dash-dotted blue line is the conditional mean normalized density weighted flame displacement speed, $\langle\widetilde{S_d}|_{\kappa}\rangle/S_L$ obtained from the 3D DNS.}
\label{fig:JPDF_Sd_kappa}
\end{figure}

\begin{figure}[h!]
\centering\includegraphics[trim=2.15cm 21.8cm 2.2cm 2.4cm,clip,width=1.0\textwidth]{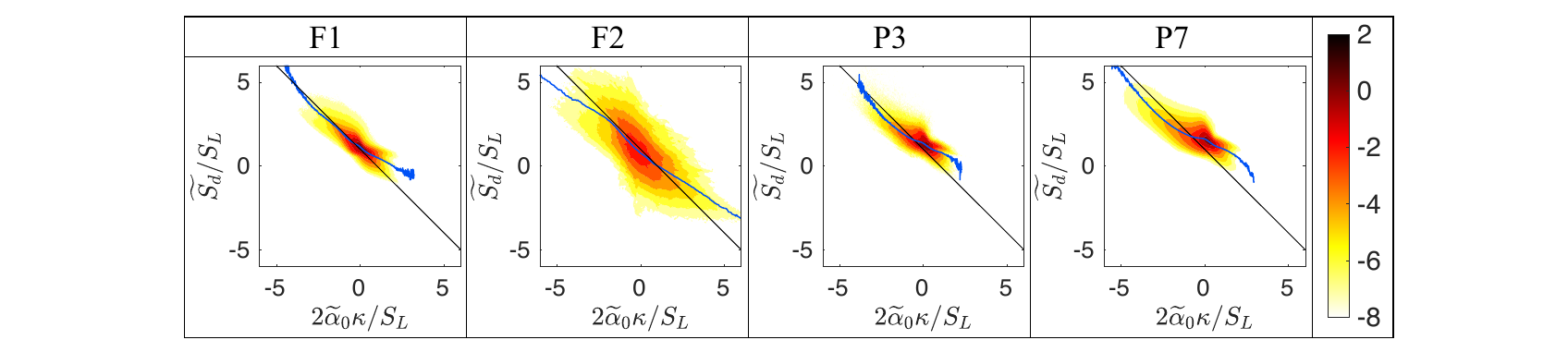}
\vspace{10 pt}
\caption{Joint probability density function (JPDF) of normalized density weighted flame displacement speed, $\widetilde{S_d}/S_L$ and 2$\widetilde{\alpha_0}\kappa/ S_L$ for cases F1, F2, P3 and P7. Each JPDF includes all the data points from all the iso-scalar surface ($c_0 =$ 0.05, 0.2, 0.4 and 0.6) put together. Colorscale represents the natural logarithm of JPDF magnitudes. The black line represents analytical results given by Eq.~(\ref{eq: final_density-weighted_Sd_kappa}). The solid blue line is the conditional mean of $\widetilde{S_d}/S_L$ for given 2$\widetilde{\alpha_0}\kappa/ S_L$, i.e. $\langle\widetilde{S_d}|_{2\widetilde{\alpha_0}\kappa/S_L}\rangle/S_L$.}
\label{fig:JPDF_Sd_kappa_all_isotherms}
\end{figure}

This subsection compares the model predictions with the direct computation of $\widetilde{S_d}$ from the DNS data at atmospheric and elevated pressure conditions. Figure~\ref{fig:JPDF_Sd_kappa} shows the joint probability density function (JPDF) of normalized density-weighted local flame displacement speed $\widetilde{S_d}/S_L$ and normalized curvature $\kappa\delta_L$ constructed from the iso-surfaces $c_0 = 0.05, 0.2,0.4,0.6$, for the four DNS cases over multiple time instances. Each of the sixteen sub-figures contains i) A solid black line representing the theoretical interaction model given by Eq.~(\ref{eq: final_density-weighted_Sd_kappa}). ii) A solid black curve with a circular marker obtained from the 1D cases of the cylindrical imploding flame, i.e., P1C, P3C and P7C. iii) Conditional mean of normalized density weighted flame displacement speed, $\langle\widetilde{S_d}|_{\kappa}\rangle/S_L$ is shown in a solid blue curve. The colorbar represents the magnitude of the natural logarithm of the probability density. The thermal diffusivity $\alpha_0$ at different $c_0$ used in Eq.~(\ref{eq: final_density-weighted_Sd_kappa}) is obtained from Chemkin-PREMIX simulation of a 1D planar laminar premixed flame. It is observed that the $\widetilde{S_d}$ is enhanced to multiple folds at large negative $\kappa$ at atmospheric and elevated pressure due to flame-flame interaction. While Eq.~(\ref{eq: final_density-weighted_Sd_kappa}) captures the trend of the 1D imploding flame very well at large negative curvatures, it explains the nature of the JPDF at all values of $\kappa$ (including $\widetilde{S_d}$ at $\kappa$=0).
Eq.~(\ref{eq: final_density-weighted_Sd_kappa}) almost overlaps with the $\langle\widetilde{S_d}|_{\kappa}\rangle/S_L$ at most $c_0$ values for the atmospheric pressure cases F1 and F2. However, at elevated pressure Eq.~(\ref{eq: final_density-weighted_Sd_kappa}) captures the $\langle\widetilde{S_d}|_{\kappa}\rangle/S_L$ fairly well on a qualitative level. The model (Eq.~(\ref{eq: final_density-weighted_Sd_kappa})) is successful in capturing the decreasing average slope of the JPDF and that obtained from 1D cases with increasing $c_0$. This occurs due to increasing density-weighted thermal diffusivity of the mixture at higher isotherms. For a given $c_0$ with increasing pressure, a decrease (increasing magnitude) in the average slope of the JPDF is observed. The model successfully captures this trend too. The magnitude of the slope ($2\widetilde{\alpha_0}/\delta_L S_L$) in Eq.~\ref{eq: final_density-weighted_Sd_kappa} is governed the density-weighted thermal diffusivity, $\widetilde{\alpha_0}$ and the normalizing factor, $\delta_L S_L$. At a given $c_0$ with increasing pressure, the drop in $\delta_LS_L$ is relatively more compared to that of $\widetilde{\alpha_0}$, leading to an overall increase in magnitude of the slope. 
Moreover, the cylindrical model is a reasonable assumption only at large negative curvatures (Fig.~\ref{fig:JPDF_kappa1_kappa2}); hence its further discussion at smaller values of $\kappa\delta_L$ is beyond the scope of the current paper.
Figure~\ref{fig:JPDF_Sd_kappa_all_isotherms} shows the JPDF of $\widetilde{S_d}/S_L$ and 2$\widetilde{\alpha_0}\kappa/S_L$ constructed from the iso-surfaces $c_0 = 0.05, 0.2,0.4,0.6$, all put together for the four DNS cases.  The straight black line represents the theoretical line given by Eq.~(\ref{eq: final_density-weighted_Sd_kappa}) and has slope = -1. The blue curve on the other hand represents the conditional mean $\langle \widetilde{S_d}|_{2\widetilde{\alpha_0}\kappa}\rangle/S_L$ over four $c_0$ values for each case. In most cases, the conditional mean (blue line) remains quasi-linear, especially at large curvatures and follows the theoretical line. In the elevated pressure cases, the conditional mean line becomes slightly non-linear near $\kappa \approx 0$. Overall, Fig.~\ref{fig:JPDF_Sd_kappa_all_isotherms} indicates the interaction model's effectiveness in qualitatively explaining the DNS data for both atmospheric and elevated pressure cases. 

\subsection{Nature of normalized density weighted flame displacement speed \texorpdfstring{$\widetilde{S_d}/S_L$}{} in the non-interacting limit, \texorpdfstring{$\kappa\rightarrow0$}{}}\label{S:3.4}
\label{S:4.3}
\begin{figure}[h!]
\centering\includegraphics[trim=2.5cm 12.4cm 2.5cm 2.5cm,clip,width=1.0\textwidth]{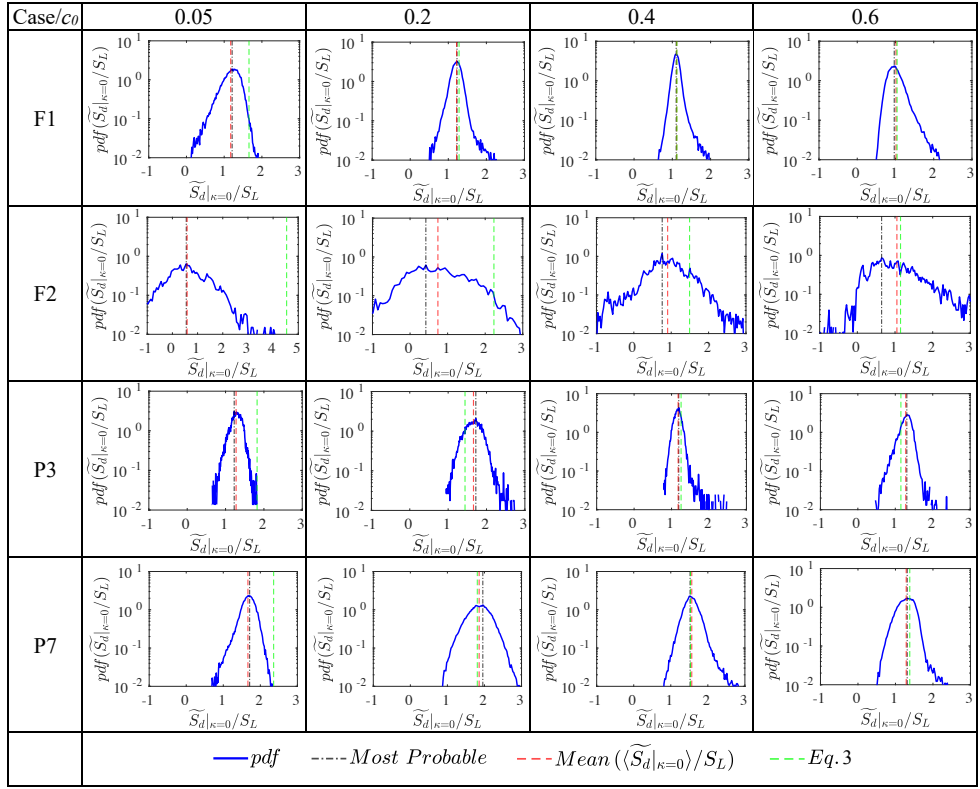}
\caption{Probability density function (PDF) of normalized density weighted flame displacement speed conditioned on $\kappa=0$, $\widetilde{S_d}|_{\kappa=0}/S_L$ (in blue) for iso-scalar, $c_0 =$ 0.05, 0.2, 0.4 and 0.6 over four-time instants for all cases. The dash-dotted black line and the dashed red line represent the most probable value and the mean of the PDF. The prediction of $\widetilde{S_d}|_{\kappa=0}/S_L$ by Eq.~(\ref{eq: Sd_eq_3}) is shown in dashed green line.}
\label{fig: PDF_SD_SL_kappa_zero}
\end{figure}

\begin{figure}[h!]
\centering\includegraphics[trim=2.5cm 13.4cm 2.5cm 2.5cm,clip,width=1.0\textwidth]{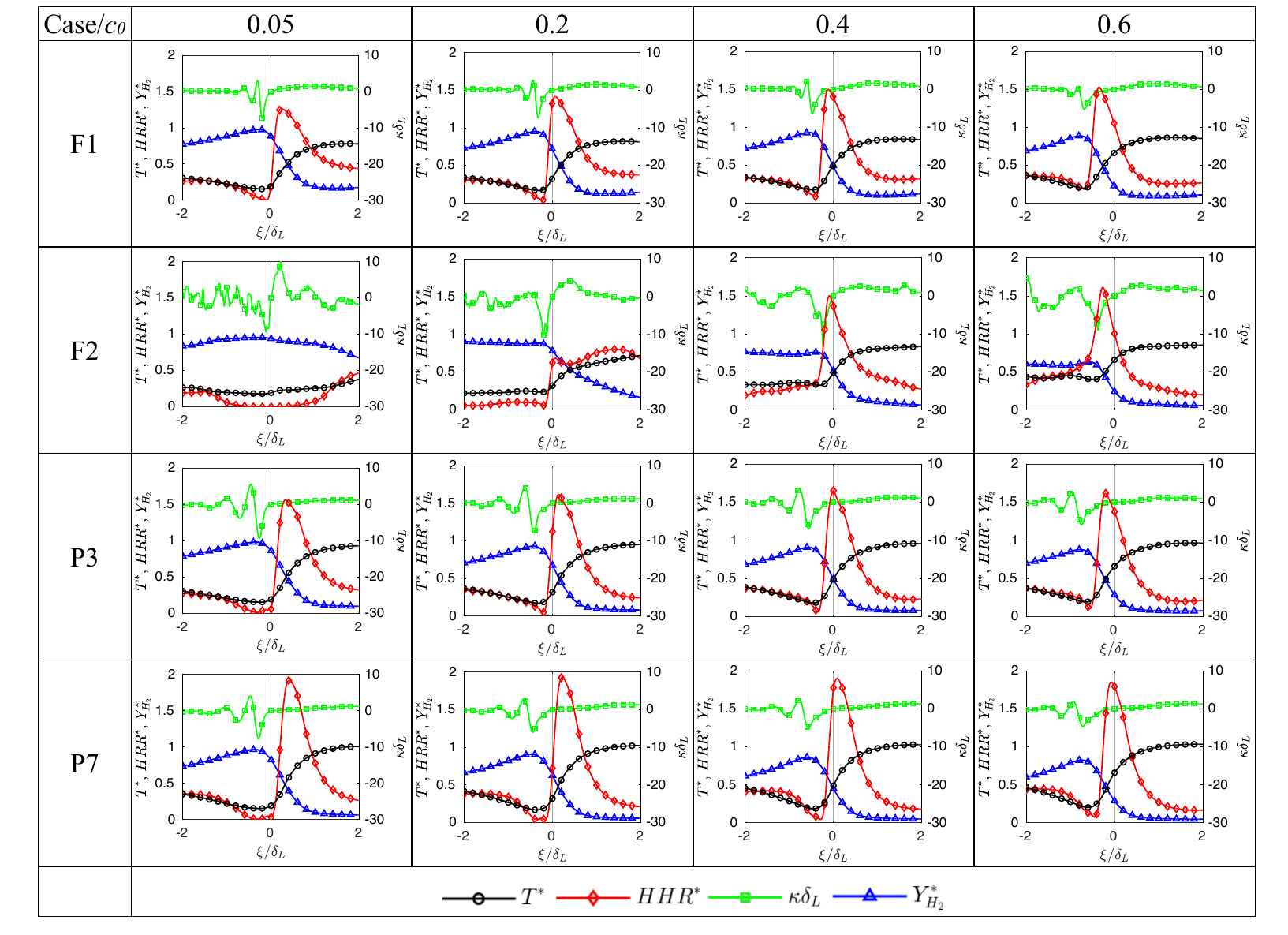}
\caption{Mean local thermo-chemical flame structures conditioned on $\kappa=0$ for iso-scalar, $c_0=0.05, 0.2, 0.4$ and $0.6$ for all cases. The faint vertical grey line denotes the point lying on the flame surface.}
\label{fig:Mean_flame_structure}
\end{figure}

It has been previously mentioned that for $\kappa=0$, we defined $\widetilde{S_d}=\widetilde{S_{d,0}}$. Hence, we investigate the probability density function (PDF) of normalized  density-weighted flame displacement speed conditioned on $\kappa=0$, $\widetilde{S_d}|_{\kappa=0}/S_L$ in Fig.~\ref{fig: PDF_SD_SL_kappa_zero}. The dash-dotted black and dashed red lines represent the most probable value and the mean value of  the PDF respectively. It should be noted that unlike $\widetilde{S_d}|_{\kappa=0}$, $\widetilde{S_{d,0}}$ can have only value which when used in Eq.~(\ref{eq: final_density-weighted_Sd_kappa_S_L*}) will best explain the JPDFs in Fig.~\ref{fig:JPDF_Sd_kappa}. Henceforth, we refer to the mean of $\widetilde{S_d}|_{\kappa=0}$ as $\widetilde{S_{d,0}}$. Figure~\ref{fig: PDF_SD_SL_kappa_zero} also shows the prediction based on the weak stretch theory (Eq.~(\ref{eq: Sd_eq_3})) in the dashed green line. The most probable values of $\widetilde{S_d}|_{\kappa=0}/S_L$ are also very close to the mean i.e. $\widetilde{S_{d,0}}/S_L$ values  for the moderately turbulent cases F1, P3 and P7 at all $c_0$ values. Perceivable differences between the means and the most probable values are observed for F2 ($c_0=0.2, 0.4 $ and $0.6$) owing to the wide scatter in the values of $\widetilde{S_d}|_{\kappa=0}/S_L$ at high $Ka$. For moderately turbulent cases (F1, P3 and P7) $\widetilde{S_{d,0}}$ shows a positive deviation from $S_L$ that increases with pressure for a given $c_0$. F2 on the other hand has $\widetilde{S_{d,0}}$ less than $S_L$ at all $c_0$ except $c_0=0.6$. For all the cases as $c_0$ increases, $\widetilde{S_{d,0}}$ eventually approaches $S_L$. Eq.~(\ref{eq: Sd_eq_3}) (dashed green line) is quite inconsistent as it only works for moderate $Ka$ cases at all pressures for higher $c_0$ but shows some discrepancies at $c_0=0.05$. Moreover, the weak stretch model fails altogether for F2. 
For further investigation, we consider the mean local thermal and chemical flame structures conditioned on $\kappa=0$ for the flame surfaces $c_0 = 0.05, 0.2, 0.4$ and $0.6$ presented in Fig.~\ref{fig:Mean_flame_structure}. At first, temperature, heat release rate, curvature and mass fraction of hydrogen are interpolated on a finite number of query points along the normal $\boldsymbol{n}$ to the flame surface, drawn at locations conditioned on $\kappa=0$. The interpolation is performed on these query points ranging from $-2\delta_L$ in the $-\boldsymbol{n}$ direction to $+2\delta_L$ in the $+\boldsymbol{n}$ direction with the origin located on the flame surface ($\xi/\delta_L=0$). Once the local flame structures are obtained, averaging is performed along the flame surface for temperature, mass fraction, curvature and heat release values at each of the query points lying on the discretized local normal to obtain the mean local thermo-chemical flame structure for that flame surface (Fig.~\ref{fig:Mean_flame_structure}). The detailed algorithm for extraction of the local flame structures and their average has been discussed by \citet{yuvraj2022local}. In Fig.~\ref{fig:Mean_flame_structure}, $T^*$, $HRR^*$, and $Y_{H_2}^*$ are the temperature, heat release rate and mass fraction of H$_2$ normalized by their corresponding maximum laminar values at that $c_0$. $\kappa\delta_L$ is the normalized curvature. The faint vertical grey line at $\xi/\delta_L=0$ denotes the point on the flame surface around which the local flame structures are extracted. The mean local flame structure brings forth two major revelations. Firstly, for the atmospheric pressure cases F1 and F2 (excluding $c_0 = 0.05, 0.2$) with $Le_{eff} = 0.758$, the heat release rate is enhanced and peaks at around 1.3-1.5 times its corresponding laminar values. Moving onto higher pressure cases $HRR^*$ increases beyond 1.5 for P3 ($Le_{eff} = 0.698$) and  for P7 ($Le_{eff} = 0.656$) it is approximately 2. This leads to the possible increase in the temperature gradient and hence $\widetilde{S_{d,0}}>S_L$. Secondly, for F2 given that the turbulence intensity is very high ($Ka\sim \mathcal{O}(1000)$), the mean flame structure is completely disrupted at lower $c_0$ values and the preheat zone is broadened. Decreased temperature gradient results in $\widetilde{S_{d,0}}<S_L$. Moving to higher $c_0$, we observe that as the turbulence is dissipated the flame structure is recovered with $\widetilde{S_{d,0}}$ eventually approaching $S_L$. Given the high $Ka$, the stretch rate is large for F2, but its effect on the flame surface is short-lived. The transient nature of the turbulence attenuates the flame speed response to the stretch rate \cite{chu2022,im2000effects,suillaud2022} and hence, the predictions of $\widetilde{S_{d,0}}$ based on the weak stretch model (Eq.~(\ref{eq: Sd_eq_3})) is proved incorrect for F2. Moreover, in intense turbulence diffusion coupled with the stretching history may also affect the $\widetilde{S_{d,0}}$ \cite{villermaux2019}. Thus it is evident that $\widetilde{S_{d,0}}$ is associated with the local flame structure, which is governed by the interplay between the turbulence intensity and the stretch rate.

\begin{figure}[h!]
\centering\includegraphics[trim=5.5cm 21cm 3cm 2cm,clip,width=1.2\textwidth]{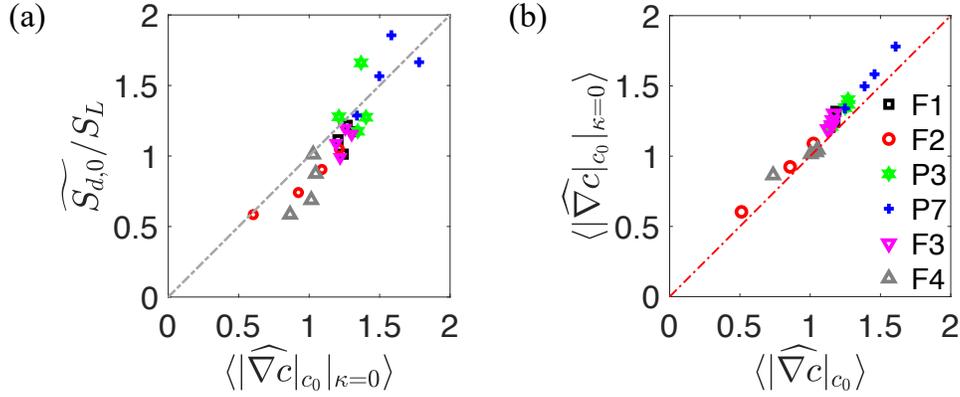}

 \caption{(a) Normalized density weighted flame displacement speed conditioned on zero local curvature ($\kappa=0$), $\widetilde{S_{d,0}}/S_L$ vs normalized averaged gradient of the progress variable conditioned on zero local curvature ($\kappa=0$), $\langle|\widehat{\nabla c}|_{c_0}|_{\kappa=0}\rangle$. The dash-dotted grey line represents the relation given by Eq.~(\ref{Ka_Le-S_L}). (b) $\langle|\widehat{\nabla c}|_{c_0}|_{\kappa=0}\rangle$ vs normalized averaged gradient of the progress variable, $\langle|\widehat{\nabla c}|_{c_0}\rangle$. The dash-dotted red line has a eslope.}
\label{fig:SL_star_gradT}
\end{figure}

Clearly, a model for $\widetilde{S_{d,0}}$ is desired. At the fundamental level $S_d$ is a representation of the local flame structure. For a laminar flame, stretch rate, curvature alters the local flame structure when $Le \neq 1$ and that is reflected as altered flame speed. However, with highly transient stretch rates in turbulence the effect of local stretch rates on the flame structure is non-trivial to capture. Hence, here we explore the possibility of correlating $\widetilde{S_{d,0}}$ with a variable that captures the variations in local flame structure, directly. The mean gradient of the progress variable on each iso-scalar can be a representation of the local flame structure. Here, we present the scatter plot of $\widetilde{S_{d,0}}/S_L$ and normalized averaged gradient of the progress variable conditioned on zero local curvature, $\langle|\widehat{\nabla c}|_{c_0}|_{\kappa=0}\rangle$ for the cases F1, F2, P3, P7, F3 and F4 in Fig~\ref{fig:SL_star_gradT}a. The cases F3 and F4 have been taken from the previous work by \cite{song2020dns, yuvraj2022local}.
 Prior to discussing the corresponding results, we revisit a planar stretched laminar flame discussed in detail in \citet{law2006}, as a well understood reference case to evaluate the dependence of the $\widetilde{S_{d,0}}$ on the $\langle|\widehat{\nabla c}|_{c_0}\rangle$ i.e., the flame structure. 

We consider quasi-one-dimensional flow through the stationary flame structure. The area of the streamtube varies owing to the stretch. For detailed configuration the reader can refer to \citet{law2006}. The governing equations are then  solved separately for the preheat and the reaction layers. Here, $f=\rho u$ is the local mass flux and $l_d = (T_b-T_u)/|dT/dx|_{x^-_f}$ with $|dT/dx|_{x^-_f}$ being the temperature gradient at the boundary of the preheat zone and the reaction layer. The subscript $u$ and $b$ refer to the unburned state and burned state whereas ``$\circ$"  denote the corresponding parameters for an unstretched planar laminar flame.

Following the analysis by \cite{law2006} we get,

\begin{equation}\label{flux-thickness_equidiffusion}
    f_u l_d=f_u^\circ l_d^\circ=\frac{\lambda_b}{C_p} 
\end{equation}

\begin{equation}\label{flux-thickness_overall}
    \widehat{f_u}= (\widehat{l_d})^{-1}
\end{equation}

\noindent where, $\widehat{f_u}=f_u/f_u^\circ$ and $\widehat{l_d}=l_d/l_d^\circ$. Assuming a linear temperature profile \cite{law2006} we have, $|dT/dx|_{x^-_f} = |dT/dx|_{T_0}= (T_b^\circ-T_u)|dc/dx|_{c_0}$, we write,

\begin{equation}\label{delta_L_tilda}
    (\widehat{l_d})^{-1}= \frac{T_b^\circ-T_u}{T_b-T_u}\frac{|\nabla T|_{T_0}}{|\nabla T|_{T_0}^\circ}
\end{equation}

\noindent with $T_b^\circ$ being the adiabatic flame temperature \cite{law2006}. The subscript $x_f^-$ denotes the boundary between the preheat and reaction zone. In addition, the mass flux can be rewritten as $f_u=\rho_uS_{L,\mathbb{K}}$ for the stretched one-dimensional laminar flames. Thus, using Eq.~(\ref{flux-thickness_overall}) and the definition for the mass flux in Eq.~(\ref{delta_L_tilda}), we get,

\begin{equation}\label{S_L-gard-T}
\frac{S_{L,\mathbb{K}}}{S_L}= \frac{1}{\widehat{T_b}-\widehat{T_u}}\frac{|\nabla c|_{c_0}}{|\nabla c|_{c_0}^\circ}
\end{equation}

\noindent where $\widehat{T}=C_pT/qY_u$ is the non-dimensional temperature and $C_p$ is considered constant. Without any heat loss, the total heat release is consumed to increase the temperature of the reactants up to the adiabatic temperature, i.e., $C_p(T_b^\circ-T_u)=qY_u$ for an unstretched planar laminar flame. For a stretched flame $\widehat{T_b}$ can be approximated as $\widehat{T_b}=\widehat{T_b^\circ} + \epsilon$ where $\epsilon=\big(\frac{1}{Le}-1\big)\mathcal{K}$, $\mathcal{K}$ being the non-dimensional stretch rate \cite{law2006}. Since $\widehat{T_b^\circ} \! \gg \! \epsilon$, $\widehat{T_b}\approx\widehat{T_b^\circ} = \widehat{T_u} +1$ in the above equation. Writing the final form of Eq.~(\ref{S_L-gard-T}) as

\begin{equation}\label{Ka_Le-S_L}
   \frac{S_{L,\mathbb{K}}}{S_L} \approx |\widehat{\nabla c}|_{c_0}
\end{equation}

Eq.~(\ref{Ka_Le-S_L}) with a unity slope is represented in a dash-dotted grey line in Fig.\ref{fig:SL_star_gradT}a.
$\widetilde{S_{d,0}}/S_L$ shows very good correlation with $\langle|\widehat{\nabla c}|_{c_0}|_{\kappa=0}\rangle$ for all cases at all $c_0$. The extreme turbulent cases F2 and F4 ($Ka > 700$) show highly broadened preheat zone with decreased normalized mean temperature gradient, $\langle|\widehat{\nabla c}|_{c_0}|_{\kappa=0}\rangle$ resulting in $\widetilde{S_{d,0}}<S_L$. P3 ($Le_{eff}=0.698$) and P7 ($Le_{eff}=0.656$) show increased temperature gradient over its laminar counterparts owing to the enhanced heat release rate as shown in Fig.~\ref{fig:Mean_flame_structure}.
The scatter of moderately turbulent cases F1, F3 at atmospheric pressure ($Le_{eff}=0.758$) agrees very well with the idealistic case of stretched planar laminar flame (dash dotted grey line representing Eq.~(\ref{Ka_Le-S_L})). Scatter plot in Fig.~\ref{fig:SL_star_gradT}b shows almost linear behavior between $\langle|\widehat{\nabla c}|_{c_0}|_{\kappa=0}\rangle$ and normalized averaged gradient of the progress variable and $\langle|\widehat{\nabla c}|_{c_0}\rangle$ for all the cases including F3 and F4 \cite{song2020dns,yuvraj2022local}. $\langle|\widehat{\nabla c}|_{c_0}\rangle \approx \langle|\widehat{\nabla c}|_{c_0}|_{\kappa=0}\rangle$ is due to the fact that the regions on the flame surfaces with $\kappa=0$ are most probable as shown by the dark red regions in Fig.~\ref{fig:JPDF_kappa1_kappa2}, \ref{fig:JPDF_Sd_kappa} and \ref{fig:JPDF_Sd_kappa_all_isotherms}. Hence, $\langle|\widehat{\nabla c}|_{c_0}|_{\kappa=0}\rangle$ contributes majorly to $\langle|\widehat{\nabla c}|_{c_0}\rangle$ for all cases at all values of $c_0$. It should be noted that Eq.~(\ref{Ka_Le-S_L}) is valid only when $\kappa\delta_L$ is in the vicinity of zero if not zero. The relation is rendered invalid under the events of global or local \cite{yuvraj2022local} flame-flame interaction at large negative curvatures. This is due to the inherently transient nature of the interacting flames. Nevertheless, the fundamental nature of Eq.~(\ref{Ka_Le-S_L}) relating the local flame displacement speed and the structure is emphasized, given that the relation is obtained using the equations of mass and energy conservation alone. 

\section{Conclusions}

Flame structures involving extreme curvatures, resulting from flame-flame and/or flame-turbulence interaction, are ubiquitous in turbulent premixed combustion. It appears from recent literature that in both small or large $Ka$ regimes, on average, the dependence of $\widetilde{S_d}$ on $\kappa$ in the large negative $\kappa$ regime is quasi-linear. Therefore, if a linear model is adopted we need to understand two parameters: the slope and the intercept. These questions were addressed in this study by developing a comprehensive theoretical analysis, validated by 1D and 3D simulations.

To model the slope, the local points on the flame surface with large local negative $\kappa$ were modeled as imploding cylindrical laminar flame based on the insights gleaned from the geometry of the flame surfaces in their principal planes as well as the past literature. To that end, 1D datasets at atmospheric and elevated pressures were also analyzed with rigorous theory to emphasize the transient nature of the interacting flames. The transient analysis enabled the effective representation of the motion of the inner reaction layer and its distinct propagation characteristics compared to the interacting isotherms during the very short-lived interaction phase. Despite the current theoretical analysis being generic, including the effect of variable density and convection, our previously proposed interaction model was recovered.

Four 3D DNS datasets at atmospheric and elevated pressure at different $Ka$ and $Re_t$ were investigated to understand the mechanism of enhanced flame displacement speed $S_d$. Detailed reaction mechanism for H$_2$-air mixture comprising of 9 species and 23 reactions by \citet{burke2012comprehensive} was used. Topological analysis revealed that large negative or positive principal curvatures leading to near cylindrical flame surfaces are highly probable compared to the spherical topology at large negative or positive curvatures. Structures involving large negative or positive principal curvatures are predominantly locally cylindrical with large reactant fluxes, inflowing or out-flowing relative to the local flame surface, leading to large deviations of $\widetilde{S_d}$ from $S_L$. As such, the present study analyzes highly curved, cylindrical, imploding, interacting, laminar premixed flames to represent highly curved flame structures in turbulence. Such configurations necessitate transient analyses, departing from the steady, weak-stretch rate theories.

When compared with an extensive DNS database, the proposed model was shown to accurately predict the slope of the JPDF of $\widetilde{S_d}$ obtained from DNS at large negative $\kappa$ for atmospheric pressure at high $Ka$ as well as elevated pressure at low $Ka$. To good accuracy, the model captured the conditional mean of $\widetilde{S_d}$ over a large range of $\kappa$ for most cases under study. 

To understand the intercept, the structure of a stretched planar laminar flame was analyzed to investigate how $\widetilde{S_d}|_{\kappa=0}$ varies with its structure denoted by its local thermal gradient $|{\nabla c}|_{c_0}$. The results from DNS showed that $\langle\widetilde{S_d}|_{\kappa=0}\rangle/S_L$ is well correlated with $\langle|\widehat{\nabla c}|_{c_0}|_{\kappa=0}\rangle$. The developed theoretical model for interacting flames is thus demonstrated as a powerful tool in describing the distributions of $S_d$, emerging from complex transient flame-flame interactions in turbulence, across different regimes and at different pressures.

\section{Acknowledgement}

This research was enabled in part by support provided by the Natural Sciences and Engineering Research Council of Canada through a Discovery Grant, the Heuckroth Distinguished Faculty Award in Aerospace Engineering from UTIAS. In addition, computational resources were provided by KAUST Supercomputing Laboratory (KSL), alongside support from KAUST. The computational resources were also provided by the SciNet High-Performance Computing Consortium at the University of Toronto and the Digital Research Alliance of Canada (the Alliance).




\bibliographystyle{elsarticle-num-names}
\bibliography{sample.bib}







\end{document}